\documentclass[12pt,preprint]{aastex}

\usepackage{epsfig}

\begin{document}

\newcommand{\vdag}{(v)^\dagger}
\newcommand{\myemail}{fukumura@physics.montana.edu}


\shorttitle{Non-Equatorial Isothermal Shocks}
\shortauthors{Fukumura et al.}

\title{Isothermal Shock Formation in Non-Equatorial Accretion Flows
around Kerr Black Holes}

\author{Keigo Fukumura\footnote{http://www.physics.montana.edu/students/keigo/homepage/html/research.html}
and Sachiko Tsuruta} \affil{Department of Physics, EPS Building,
Montana State University, Bozeman, MT 59717-3840; \myemail,
tsuruta@physics.montana.edu}

\begin{abstract}
We explore isothermal shock formation in non-equatorial, adiabatic
accretion flows onto a rotating black hole, with possible
application to some active galactic nuclei (AGNs). The isothermal
shock jump conditions as well as the regularity condition,
previously developed for one-dimensional (1D) flows in the
equatorial plane, are extended to two-dimensional (2D),
non-equatorial flows, to explore possible geometrical effects. The
basic hydrodynamic equations with these conditions are
self-consistently solved in the context of general relativity to
explore the formation of stable isothermal shocks. We find that
strong shocks are formed in various locations above the equatorial
plane, especially around a rapidly-rotating black hole with the
prograde flows (rather than a Schwarzschild black hole). The
retrograde flows are generally found to develop weaker shocks. The
energy dissipation across the shock in the hot non-equatorial
flows above the cooler accretion disk may offer an attractive
illuminating source for the reprocessed features, such as the iron
fluorescence lines, which are often observed in some AGNs.
\end{abstract}


\section{Introduction}

A realistic model of the AGN central engine may include the
effects of the magnetic field. The black hole magnetosphere is
studied first by \cite{BZ77} in the context of winds and jets from
radio-loud AGNs. The work has been extended to magnetospheric
physics of accreting AGNs by various authors
\citep[e.g.,][hereafter TRFT02; Rilett et al.2004, in
preparation]{Phinney83,TNTT90,TT01,TRFT02}. In this case, plasma
particles should be {\it frozen-in} to the magnetic field lines,
and hence the accreting fluid should fall onto the black hole from
regions {\it above} the equatorial plane along the field
lines\citep[see, e.g., Figure 2 of TRFT02; Figure 1 of][]{TT01}.
Then, magnetohydrodynamics (MHD) should become important to
describe the motion of the particles associated with the
background field. The resulting relativistic MHD shocks are
explored by TRFT02 and Rilett et al. (2004, in preparation).

In general, the MHD shocks can be hydro-dominated or
magneto-dominated \citep[hereafter T02]{Takahashi00,Takahashi02}.
Obviously the MHD case is, however, very complicated. Therefore,
the main motivation of our current paper is, as a starting point,
to investigate the hydrodynamic limit, which should be valid in
the case of small magnetization. Here, we adopt a model which
should apply to the hydro-dominated shocks where the magnetic
field does not make a significant contribution to the properties
of the shocks under a weak field limit, because in such a case
hydrodynamics should primarily control the shock formation.

One-dimensional (1D), hot accretion flows around a black hole,
generally treated as ideal hydrodynamical fluid, have been
investigated by various authors \citep{Sponholz94,Kato96,Kato98}.
It has been found that such accretion flows must be transonic, and
will become supersonic before reaching the event horizon, while it
is subsonic at infinity. Once such a fluid becomes supersonic, it
is likely that a standing shock wave will develop when shock
conditions are met. There are roughly three types of standing
shocks: adiabatic (Rankine-Hugoniot) shocks, isothermal shocks,
and isentropic compression waves \citep{Abramowicz90}. The first
attempt was made by \cite{Yang95} to self-consistently study the
relativistic isothermal shock formation around black holes. In the
case of isothermal shocks, the postshock fluid can lose
substantial energy and entropy across the shock while the fluid
temperature is continuous across the shock location \citep{Lu97b}.
\citet[hereafter LY98]{Lu98} examined the isothermal shock
formation in one-dimensional (1D) adiabatic hot flows in the Kerr
geometry, for various flow parameters including black hole
rotation.

In our current paper, we extend the work by LY98 on isothermal
shock formation in 1D adiabatic flows in the equatorial plane, to
two-dimensional (2D) calculations for flows {\it above} the
equatorial plane, to investigate geometrical effects. One of our
major motivations is to explore the possibility that shocks
produced in such flows act as a high energy radiation source for
some reprocessed features, such as the iron fluorescent lines,
which are observed from some AGNs. It is generally considered that
a hot illuminating source {\it above} the cooler disk is required
to produce such features \citep{Fab89}.

Section \S{2} introduces our basic equations and assumptions. The
hydrodynamic fluid equations are solved in the context of general
relativity. We discuss the isothermal shock conditions and the
stability of shocks. The results are presented in Section \S{3}
where we display shocks for various representative values of
angular position, fluid energy, angular momentum, and black hole
spin. Discussion and concluding remarks are given in the last
section, \S{4}.

\placefigure{fig:disk}

\section{Basic Equations \& Assumptions}

\subsection{Background Kerr Geometry}
The space-time is assumed to be stationary ($\partial/\partial
t=0$) and axially-symmetric ($\partial/\partial \phi=0$) around a
rotating black hole. The background Kerr metric is then expressed
in the Boyer-Lindquist coordinates as
\begin{eqnarray}
   ds^2 &=& -\left( 1-\frac{2Mr}{\Sigma} \right) dt^2
        - \frac{4aMr\sin^2\theta}{\Sigma} \,dt d\phi \nonumber \\
        & & + \frac{A\, \sin^2\theta}{\Sigma} \, d\phi^2
        + \frac{\Sigma}{\Delta}\, dr^2 + \Sigma\, d\theta^2 \ ,
\end{eqnarray}
where $\Delta \equiv r^2 -2Mr +a^2 $, $\Sigma \equiv
r^2+a^2\cos^2\theta$, $A \equiv (r^2+a^2)^2-a^2\Delta
\sin^2\theta$ with the metric signature being $(-,+,+,+)$.
Throughout this paper, the distance is normalized by the
gravitational radius $r_g \equiv GM/c^2$ where $c,G,M$ being the
speed of light, gravitational constant, and the black hole mass,
respectively. The hole's event horizon is then $r_h \equiv
(1+\sqrt{1-a^2}) r_g$. Notice that $\theta = \theta_{sh} = \pi /2$
corresponds to the shocks in the equatorial plane while $\theta =
\theta_{sh} \neq \pi /2$ denotes the non-equatorial shocks
discussed here.

\subsection{Hot, Adiabatic Fluid above the Equatorial Plane}
\begin{figure}
    \epsscale{0.3}
    \plotone{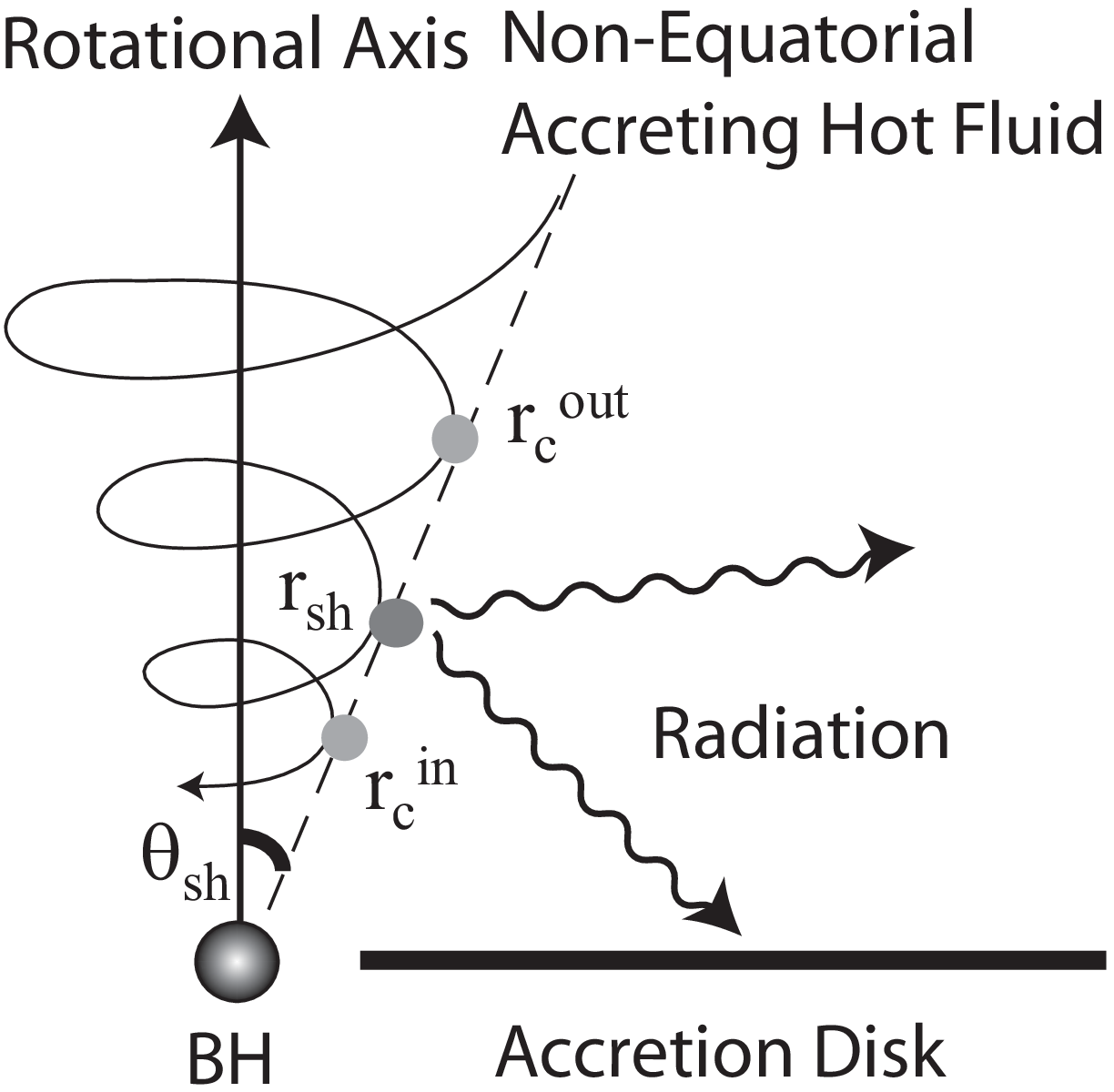}
    \caption{A side view of the schematic geometry in our model. A non-equatorial accreting fluid
             is conically accreting in the poloidal plane ($u^{\theta}=0$) onto the black hole, spiralling around the
             rotational axis
             $(u^r<0, u^{\phi} \neq 0)$ with a constant angle
             $\theta_{sh}$. It passes through
             the outer critical point $r^{out}_c$, gets shocked at $r_{sh}$ and passes through the inner critical point
             $r^{in}_c$ before reaching the horizon. The shock location $r_{sh}$ increases with increasing specific angular
             momentum of the fluid $\ell$. See the text for details.}
    \label{fig:cross-section}
\end{figure}

\cite{Cha96a,Cha96b}, \cite{Lu97a} and LY98 have studied general
relativistic, equatorial flows in the Kerr geometry, whereas many
of the previous works were based on the Pseudo-Newtonian potential
\citep{Paczynski80}, which is not capable of reproducing the
frame-dragging effect. To examine the two-dimensional (2D),
relativistic hot accretion flow, here we assume an ideal,
Boltzmann gas in a non-equatorial plane. Such an accreting fluid
spirals around the symmetry axis due to angular momentum and
accretes onto the black hole. The poloidal path of the fluid is
assumed to be conical $(u^{\theta}=0)$ since the preshock fluid
slowly starts falling from a distant location as a subsonic flow.
That is, the fluid has only the radial velocity $(u^r<0)$ and the
azimuthal velocity $(u^{\phi} \neq 0)$.  Therefore, flow particles
will spiral around the black hole rotation axis onto the hole. The
schematic diagram of such an accreting fluid is displayed in
Figure~\ref{fig:cross-section}.  The non-equatorial accreting
fluid is conically falling onto the hole (i.e., $\theta$=constant
and $u^r<0$ in the poloidal plane), spiralling around the symmetry
axis in the azimuthal plane.

\placefigure{fig:cross-section}

In this model perturbing forces to the fluid that might cause any
acceleration in the $\theta$ direction is assumed to be
negligible, keeping the flow at constant $\theta$ in a conical
flow. This assumption is justified because we are exploring the
weak field limit of MHD flows, which should apply to flows with
small magnetization. As noted in Section \S{1} already, fluid
particles in such flows should be frozen-in to the field lines and
flow along the filed lines. Therefore, the flow geometry in our
case is governed by the magnetospheric structure, rather than by
the hydrostatic equilibrium of non-magnetic fluid. In this sense
it is not appropriate, in our case, to adopt the approach of
conventional, non-magnetic 2D thick accretion disk (or torus)
models. Especially, it may be emphasized that our version of the
`conical model' is quite different from the conical equilibrium
flow often adopted in conventional non-magnetic thick disk models.

Solving the geometry of the magnetospheric structure in a
realistic manner is an extremely difficult problem, and only a
very simple approximate version has been carried out (see, e.g.,
Tomimatsu and Takahashi 2001). Therefore, in our current paper
which should apply to the weak field limit, we use, as our first
approximation, the flow geometry we adopted already in our
previous MHD accretions model where the infalling plasma is
frozen-in to the radial field lines in a conical geometry, and
hence the flows in different $\theta$ directions are decoupled
from each other (see TRFT02, Rilett et al. 2004, in preparation).

Following the earlier works on relativistic accretion shocks,
particularly LY98, we assume that a dynamical time-scale of the
accretion process is much shorter than that for the energy (or
thermal) dissipation during the fluid accretion. The fluid obeys
the equation of state for an ideal gas
\begin{eqnarray}
P = {k_B \over \mu_p m_p} \rho T \ , \label{eq:EOS}
\end{eqnarray}
where the primary thermodynamic property of the fluid is
characterized by the locally measured temperature $T$ and the
thermal pressure $P$. $\rho$, $\mu_p$ and $m_p$ are the rest-mass
density of the fluid, the mean molecular weight of the composite
particles and the mass of a particle, respectively. $k_B$ is the
Boltzmann constant. The polytropic form is adopted as
\begin{eqnarray}
P &=& K \rho^{\gamma} = K \rho^{1+1/n} \ , \label{eq:polytrope}
\end{eqnarray}
where the adiabatic index $\gamma$ is constant, and the polytropic
index $n$ is correspondingly constant, too. We use $\gamma=4/3$
(or $n=3$) for our relativistic fluid. $K$ is a measurement of the
entropy $S$ of the gas where $S \equiv c_v \log K$ ($c_v$ being a
specific volume heat), expressed also via equations (\ref{eq:EOS})
and (\ref{eq:polytrope}) as
\begin{eqnarray}
K &\equiv& {k_B \over \mu_p m_p \rho^{\gamma-1}} T \ .
\end{eqnarray}
We consider the fluid to be stationary and axisymmetric. The fluid
trajectory is conical in the poloidal plane and is spiralling
around the polar axis onto the hole. Thus, $\theta$=constant or
$u^{\theta}=0$, but $u^r<0$ and $u^{\phi} \neq 0$. Due to the
space-time symmetry, specific total energy $E$ and axial angular
momentum component $L$ of the fluid are conserved along the
shock-free fluid's path and are written as
\begin{eqnarray}
E &\equiv& -\mu u_t  \label{eq:E} \ ,  \\
L &\equiv& \mu u_{\phi} \ ,
\end{eqnarray}
where $\mu=(P+\epsilon)/\rho$ is the relativistic enthalpy of the
fluid, and $\epsilon=\rho + nP$ is the total energy-density
including the internal energy term. With the use of the
four-velocity normalization {\it $\textbf{u}$} $\cdot$ {\it
$\textbf{u}$} $= -1$ where {\it
\textbf{u}}~=~$(u^t,u^r,0,u^{\phi})$ is the four-velocity of the
fluid, we find
\begin{eqnarray}
1 + u_r u^r + (u^t)^2 V_{eff}(r,\ell) = 0 \ ,
\end{eqnarray}
where $V_{eff}(r,\ell) \equiv g^{tt}-2 \ell g^{t\phi}+\ell^2
g^{\phi\phi}$ is called the effective potential \citep{Lu95} with
$g^{\alpha \beta}$ being the Kerr metric tensor components. $\ell
\equiv L/E = -u_{\phi}/u_t$ is the specific angular momentum of
the fluid which is conserved along the whole flow if viscous
dissipation is negligibly small (weak viscosity limit). Hence, we
get $u_t$ in terms of $V_{eff}(r,\ell)$ as
\begin{eqnarray}
u_t = \left(\frac{1+u_r u^r}{-V_{eff}} \right)^{1/2} \ .
\end{eqnarray}
On the other hand, the definition of the local sound speed is
given by $c_s \equiv (dP / d \epsilon)^{1/2}$ following LY98,
which is then rewritten as
\begin{eqnarray}
c^2_s \equiv \frac{d P}{d \epsilon} = \left(1+\frac{1}{n} \right)
K \rho^{1/n} \left(\frac{d \rho}{d \epsilon} \right)
\label{eq:sound} \ ,
\end{eqnarray}
and by definition of the enthalpy we also have
\begin{eqnarray}
\mu = \frac{P+\epsilon}{\rho} = 1+(1+n) K \rho^{1/n}
\label{eq:enthalpy} \ .
\end{eqnarray}
Thus, combining equations~(\ref{eq:sound}) and
(\ref{eq:enthalpy}), we find
\begin{eqnarray}
1-n c^2_s = 1-n \left(\frac{d P}{d \epsilon} \right) =
\frac{1}{1+(1+n)K \rho^{1/n}} = \frac{1}{\mu} \label{eq:enthalpy2}
\ ,
\end{eqnarray}
where we now express the local sound speed in terms of the
enthalpy (or vice versa). Accordingly, therefore,
equation~(\ref{eq:E}) can be rewritten as
\begin{eqnarray}
E = \left({1+u^ru_r \over -V_{eff}} \right)^{1/2} / \left(1-n
c^2_s \right) \label{eq:energy} \ .
\end{eqnarray}
It is useful to notice here that both $E$ and $L$ will change
across the shock location in the case of isothermal shock, but
they change in such a way that the ratio $L/E$ (i.e., $\ell$) will
stay unchanged (see a subsequent section). Notice that $K$ remains
the same along the shock-free fluid due to the adiabatic
assumption (no heat in/out), but does change across the shock.
There is another constant along the fluid's path called mass
accretion rate $\dot{M}$. The spherical mass accretion rate is $4
\pi r^2 (-u^r) \rho$ in the radial direction, and thus the mass
accretion rate per unit spherical area by incoming fluid is given
by
\begin{eqnarray}
\dot{M} = r^2 (-u^r) \rho \ ,
\end{eqnarray}
which is constant along the whole flow. Since we are interested in
shock formation where entropy is generated, let us follow LY98 and
introduce another conserved quantity called the entropy accretion
rate
\begin{eqnarray}
\dot{\cal{M}} \equiv K^n \dot{M} = r^2 \left[{c^2_s \over \gamma
(1-nc^2_s)} \right]^n (-u^r) \label{eq:Mdot} \ ,
\end{eqnarray}
where $\dot{\cal{M}}$ is conserved in a shock-free fluid, but
changes in a shock-included fluid due to the entropy generation
(i.e., $K$) at the shock location.

\subsection{Regularity Conditions}
It is known that any black hole accretion must be transonic
\citep{Cha90a,Cha96a}, so it is important to locate the critical
points (or sonic points to a particular observer) to make sure
that the fluid is physically acceptable. Taking a derivative of
equation~(\ref{eq:energy}) with respect to $r$, we obtain
\begin{eqnarray}
{du^r \over dr} = {N \over D} \label{eq:ND} \ ,
\end{eqnarray}
where
\begin{eqnarray}
N &\equiv& {(u^r)^2 \over 2 (1+u^r u_r)} {dg_{rr} \over dr} - {2
c^2_s \over r} - {1 \over 2 V_{eff}} {dV_{eff} \over dr} \ ,  \label{eq:N} \\
D &\equiv& {c^2_s \over u^r} - {u_r \over 1+u^r u_r} \ .
\label{eq:D}
\end{eqnarray}
In equation~(\ref{eq:ND}), $D(N=0)=0$ is required (regularity
conditions) for $du^r/dr$ to be finite. Therefore, at critical
points we get
\begin{eqnarray}
c^2_s|_c &=& {u_r u^r \over 1+u_r u^r}|_c \ ,  \label{eq:critical1} \\
c^2_s \left[{d(lng_{rr}) \over dr} - 4/r \right]|_c &=& {1 \over
V_{eff}} {dV_{eff} \over dr}|_c \ ,  \label{eq:critical2}
\end{eqnarray}
where the subscript ``c'' denotes the critical points. Notice that
the right-hand-side of equation~(\ref{eq:critical1}) is the radial
three-velocity component of the fluid measured by a corotating
observer (this will be explained later). In other words, the
critical points are equivalent to the sonic points only for such
an observer. Multiple critical points (i.e., an inner point
$r^{in}_c$, an intermediate point $r^{mid}_c$ and an outer point
$r^{out}_c$) are known to exist in general for a set of $(E,\ell)$
\citep[e.g.,][]{Cha90b}. This means that the location of the
critical points will be different before and after the shock in a
shock-included fluid since $E$ changes across the shock. The fluid
must pass through a critical point on both sides of the shock. It
is worth while to note that in isothermal shocks the postshock
fluid cannot pass through the same critical points for the
preshock fluid due to the energy dissipation at the shock
location. This will be addressed more in a later section.
Depending on the critical point that the fluid goes through,
kinematic and thermodynamic profiles of the fluid will be
different. In the framework of our model, however, it is not
terribly important through which critical point the accreting
fluid will pass as far as it produces a dissipative shock near the
black hole. Therefore, we will only investigate the preshock fluid
that passes through the outer critical point $r^{out}_c$ and the
postshock fluid that passes through the inner critical point
$r^{in}_c$.

\subsection{Isothermal Shock Conditions}
In our scenario the adiabatic fluid forms a standing shock (i.e.,
stationary shock location) somewhere between the preshock outer
critical point $r^{out}_c$ and the postshock inner critical point
$r^{in}_c$ on its way to the horizon. According to
\cite{Abramowicz90}, standing shocks in general can be categorized
into three types: (1)adiabatic shocks or Rankine-Hugoniot shocks,
(2)isentropic compression waves and (3)isothermal shocks. In the
case of (1), the fluid by definition does not release any energy
across the shock, carrying the generated entropy and its thermal
energy with it \citep{Fukue87,Cha90b,Lu97a}. Thus, radiative
cooling mechanism is extremely inefficient with the thermal energy
being advected. In type (2), the shock radiates an energy
equivalent to the generated entropy at the shock such that the
entropy remains unchanged at the shock \citep{Cha89}. In the case
of type (3), a fraction of the preshock fluid's energy and the
entropy generated are lost from the fluid's surface at the shock
location such that the temperature (therefore the sound speed)
remains continuous across the shock \citep[][LY98]{Lu97b}. In
other words, the dissipative cooling processes are so efficient
that the energy is not advected with the fluid. A realistic shock
could be between these two extremes (adiabatic and isothermal).
However, in this paper we investigate case (3), because it is very
interesting in the sense that a large amount of energy releases
can be expected in the vicinity of the central black hole, which
may be related to the observed emission features in AGNs.

Since the fluid total energy decreases across the isothermal
shock, we have
\begin{eqnarray}
E_1 &=& \left({1+u_1^r u_{r1} \over -V_{eff}} \right)^{1/2} /
\left(1-n c^2_{sh} \right) \ , \label{eq:energy1} \\
E_2 &=& \left({1+u_2^r u_{r2} \over -V_{eff}} \right)^{1/2} /
\left(1-n c^2_{sh} \right) \ , \label{eq:energy2} \\
E_{sh} &\equiv& E_1 - E_2 \ , \label{eq:fsh}
\end{eqnarray}
where the subscript ``1'' and ``2'' denote a preshock and
postshock quantity at the shock location, respectively. $c_{sh}
\equiv c_s(r_{sh})$ is the local sound speed at the shock
location. From the entropy accretion rate, we have
\begin{eqnarray}
\dot{\cal{M}}_1 = r^2_{sh} \left[{c^2_{sh} \over \gamma (1-n c^2_{sh})}
\right]^n (-u^r_1) \ , \label{eq:Mdot1} \\
\dot{\cal{M}}_2 = r^2_{sh} \left[{c^2_{sh} \over \gamma (1-n
c^2_{sh})} \right]^n (-u^r_2) \ . \label{eq:Mdot2}
\end{eqnarray}
Due to the shock formation, the fluid becomes hotter and entropy
is generated at the shock front (i.e., mathematical discontinuity
with zero-thickness). Thus, $\dot{\cal{M}}$ increases
instantaneously at the shock location. However, the radiative
cooling process may be so efficient that a fraction of the total
entropy may be lost as radiation from the fluid surface.
Therefore, the postshock temperature remains continuous and the
postshock entropy becomes smaller than that of the preshock. That
is, $T_1 = T_2$ but $\dot{\cal{M}}_1 > \dot{\cal{M}}_2$ although
$\dot{M_1} = \dot{M_2}$. The radial momentum flux density is given
by
\begin{eqnarray}
F \equiv {c^2_s \over \gamma u_r} + u^r.
\end{eqnarray}
From conservation of momentum flux density in the radial
direction, it is found that
\begin{eqnarray}
{c^2_{sh} \over \gamma u_{r1}} + u^r_1 = {c^2_{sh} \over \gamma
u_{r2}} + u^r_2 \ . \label{eq:momentum}
\end{eqnarray}
Based on the above jump conditions in addition to the regularity
conditions, there are 9 equations
(\ref{eq:energy}),~(\ref{eq:Mdot}),~(\ref{eq:critical1}),~(\ref{eq:critical2}),
~(\ref{eq:energy1}),~(\ref{eq:energy2}),~(\ref{eq:Mdot1}),~(\ref{eq:Mdot2})
and (\ref{eq:momentum}) for 9 unknowns $r^{in}_c$, $r^{out}_c$,
$r_{sh}$, $E_2$, $\dot{\cal{M}}_1$, $\dot{\cal{M}}_2$, $c_{sh}$,
$u^r_1$ and $u^r_2$. Once a set of ($a, \theta_{sh}, E_1, \ell$)
is specified, then the corresponding shock location $r_{sh}$ and
all the other unknown quantities are uniquely determined. For the
sake of completeness, it should be noted that any physically
acceptable shocks in a transonic accretion flow must satisfy the
boundary conditions ($r^{in}_c < r_{sh} < r^{out}_c$).

From equations~(\ref{eq:energy1}) and (\ref{eq:energy2}), we note
that $E_1>E_2$ at the shock location $r_{sh}$ because
$|u^r_1|>|u^r_2|$. That allows some energy dissipation $E_{sh}$.
Equations~(\ref{eq:Mdot1}) and (\ref{eq:Mdot2}) also imply that
the entropy accretion rate should decrease across the shock for
the same reason. Now, we may extract some interesting qualitative
implications from these equations. For example, we note that more
energy release $E_{sh}$ should naturally be expected from strong
shocks because a relatively large jump in the radial velocity
$u^r$ across such shocks is made possible. Secondly, we clearly
see the general trend for the effect of the direction,
$\theta_{sh}$, on specific angular momentum of the fluid, $\ell$.
For instance, $\ell$ cannot be large near the polar region since
the centrifugal barrier is correspondingly small compared with
that near the equator. Then, the non-equatorial fluid will have
smaller rotational velocity $u^{\phi}$. Consequently, the flow
could gain its radial component $u^r$ more efficiently than its
azimuthal component $u^{\phi}$, which allows the preshock flow to
become supersonic more quickly (or sooner) in the radial
direction. The resulting shock location, therefore, will be
further from the black hole for small $\theta_{sh}$ as compared
with the shocks near the equator (large $\theta_{sh}$).
Furthermore, the energy dissipation $E_{sh}$ from such (more
distant) shocks will be small, due to the fact that only a small
amount of fluid's gravitational potential energy can be released,
as opposed to a large amount of energy conversion from the
gravitational potential in regions close to the hole. We will come
back to this issue in Section \S{3}.

As mentioned in the earlier section, the critical points $r_c$ are
determined by the fluid parameters $(E,\ell)$. In dissipative
shock-included flows, the preshock fluid energy $E_1$ (and
$\dot{\cal{M}}_1$) decreases across the shock, which will
subsequently determine the postshock critical points corresponding
to different $E$. In other words, the postshock fluid with
$(E_2,\dot{\cal{M}}_2)$ cannot pass through the critical points
already determined by the preshock flow with
$(E_1,\dot{\cal{M}}_1)$. Instead, new critical points for the
postshock flow will be determined by the postshock flow parameter
$(E_2,\dot{\cal{M}}_2)$. Therefore, one must find these new
critical points based on $(E_2,\dot{\cal{M}}_2)$ for the postshock
flows to describe global shock-included transonic flows. Such a
flow topology will be illustrated in \S{3.5}.

\subsection{Stability Analysis}
Although possible shock locations are found from the jump
conditions, they need to be stable against a small perturbation.
Otherwise, the shock will disappear as a result of a small change
in its location. To check the stability, we perturb the radial
momentum flux density (equivalent to pressure) by a small change
in the shock location $r_{sh}$, a method originally developed by
\cite{Yang95}. Thus, a small perturbation of the momentum flux
density is
\begin{eqnarray}
\delta F_2 - \delta F_1 = \left({dF_2 \over dr} - {dF_1 \over dr}
\right) \delta r \equiv \Delta \delta r \ .
\end{eqnarray}
When the momentum flux density is not balanced on both sides of
the shock, the shock is not stable. Suppose that a shock location
is slightly perturbed by $\delta r$. If $\Delta>0$ when $\delta
r>0$ (or $\delta r<0$), the postshock momentum flux density
becomes larger (or smaller) for accretion. Consequently, the shock
location will be further shifted towards increasing (or
decreasing) r, and the fluid will never find a stable shock
location. If $\Delta<0$, on the other hand, the shock will always
go back to the original location satisfying momentum equilibrium
regardless of the sign of $\delta r$. Therefore, $\Delta<0$ is
required for stable shocks for accretion.

\subsection{Characteristics of Global, Shock-Included Flows}
It is interesting to see how flow quantities are changing as the
accreting gas falls onto the black hole. Here, we will estimate
some of the important flow quantities for global, shock-included,
transonic fluids as a function of radial distance $r$. First, we
explore the kinematics of the flows through the four-velocity
components $u^r,u^{\phi}$ and the angular velocity $\Omega \equiv
u^{\phi}/u^t$ as well. The physical flow velocity can be described
by the radial three-velocity $v^r \equiv [u_r u^r / (1+u_r
u^r)]^{1/2}$ in a corotating reference frame (CRF) which is a
locally-flat space. $v^r$ is known to be equal to the speed of
light at the horizon. The Mach number is defined in CRF by $M
\equiv v^r/c_s$. The relativistic enthalpy of the flow can be
written as $\mu \equiv 1/(1-n c^2_s)$ in the unit of fluid
rest-mass energy (see equation~(\ref{eq:enthalpy2})). The thermal
pressure $P$ can be derived from equation~(\ref{eq:polytrope}) as
$P \propto \rho^{\gamma}$. Then, the local temperature of the flow
goes as $T \propto P / \rho$ using equation~(\ref{eq:EOS}). By the
mass conservation law, we get the local density of the flow as
$\rho \propto 1/(r^2 |u^r|)$. Here, we define a normalized
pressure $\hat{P} \equiv P / K = \rho^{\gamma}$, temperature
$\hat{T} \equiv P / \rho = k_B T / (\mu_p m_p)$ and the mass
density $\hat{\rho} \equiv \rho/\dot{M} = 1/(r^2 |u^r|)$,
respectively.

\section{Results}
In this section we present our results and consider their physical
implications. The obtained isothermal shock solutions are
displayed for various parameters. Based on previous AGN
observations and theoretical speculations
\citep[e.g.,][]{Iwasawa96a,Iwasawa96b,Wilms01}, the mass of the
central supermassive black hole is chosen to be $M=10^7 M_{\sun}$,
and the hole is in general rapidly-rotating (i.e., an extreme Kerr
black hole). We choose $a=0.99$ in most cases.  However, in
Section \S{3.4} the black hole spin dependence is investigated by
adopting various values of $a$. The accretion fluid is rotating
with the Keplerian angular velocity of $\Omega_k = 1/(a+r^{3/2})$.
We consider both direct flows ($a \ell >0$) and retrograde flows
($a \ell<0$).

As seen by LY98, we find stable, multiple shock locations
($r^{in}_{sh},r^{out}_{sh}$ where $r^{in}_{sh}<r^{out}_{sh}$) for
the same set of parameters $(a,\theta_{sh},E_1,\ell)$ satisfying
all the requirements (i.e., shock jump conditions, boundary
conditions, and stability analysis). The preshock flow which does
not reach the event horizon is often called $\alpha$-type flow. In
order for $\alpha$-type preshock flows to become physically
acceptable, there must exist a shock. However, it is still
uncertain through which shock location, between $r^{in}_{sh}$ or
$r^{out}_{sh}$, the preshock flow will ``transit'' to the
postshock flow (i.e., degeneracy of shocks). There may be some
other physical conditions (for instance, viscous dissipation or
magnetic field) that uniquely determine which shock is actually
realized, but this is beyond the scope of this paper. To avoid
complexity caused by the degeneracy, we will include only the
inner range of shock locations $r^{in}_{sh}$ (which is closer to
the hole).  Although we are aware of the existence of the
``degeneracy'', throughout this paper the outer range of shock
locations are not considered. The reason is that, as mentioned
already, our work is originally motivated by the X-ray
observations of the ``reprocessed features'' such as the iron
lines from some AGNs.  These lines are considered to come from a
cool accretion disk illuminated by a hot high energy radiation
source above the disk plane. The whole system should, however, be
very close to the central black hole\citep{Fab89}, because these
lines often exhibit broadened features with a large gravitational
redshift\citep{Nandra94}.  Therefore, we are primarily interested
in the inner shock formation close to the hole.

\placefigure{fig:shock-location}

\subsection{Isothermal Shock Locations above the Equatorial Plane
($0\degr < \theta_{sh} < 90\degr$)}

\begin{figure}
    \epsscale{0.6}
    \plotone{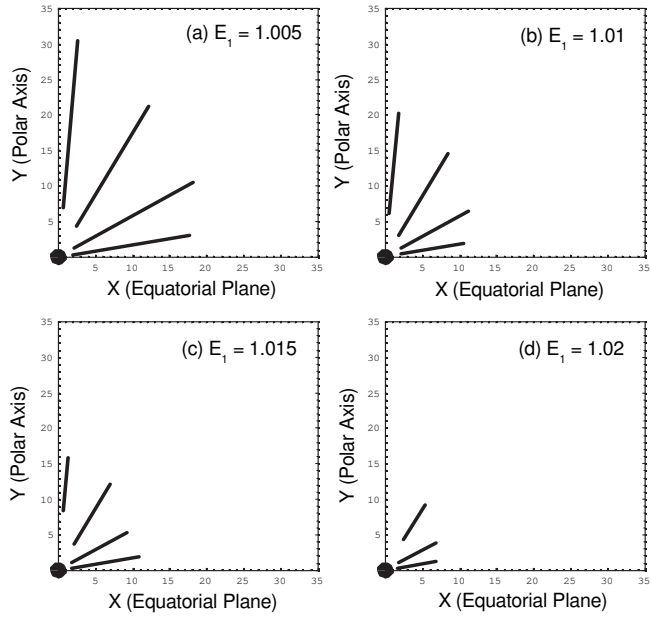}
    \caption{The range of isothermal, stable shock locations around a rapidly-rotating black hole ($a=0.99$ with prograde flows)
             for various
             angle $\theta_{sh}$ for fixed energy $E_1$. The black hole's event horizon is $r_h=1.141
             r_g$ (central dot). No shock is found for $E_1=1.02$ with
             $\theta_{sh}=5\degr$ in (d).
             Note that only the inner branch of shock locations are shown.
             }
    \label{fig:shock-location}
\end{figure}

Solving the basic equations (conservation laws + shock conditions)
introduced earlier, we found various stable shock locations for
different fluid parameters such as ($\theta_{sh}, E_1, \ell$). The
following four angles $\theta_{sh}$ are considered:
$5\degr$(near-axis region), $30\degr$(lower polar region),
$60\degr$(upper equatorial region) and $80\degr$(near equatorial
region). The energy-dependence of the shock locations for various
$E_1$ is also examined in the following manner: $E_1 = 1.005,
~1.01, ~1.015$ and $1.02$ for comparison where $E_1$ is expressed
in the unit of the particle's rest-mass energy $m_p c^2$. To avoid
complications, we fix the black hole spin at $a=0.99$ (with the
prograde flows). First, the cross-sectional distribution in the
(X,Y)-plane of the obtained shock locations is displayed in
Figure~\ref{fig:shock-location} where $E_1$ is (a)1.005, (b)1.01,
(c)1.015 and (d)1.02 for the four selected $\theta_{sh}$ :
$5\degr,~30\degr,~60\degr$ and $80\degr$ from top to bottom except
for (d). As explained earlier, the outer range of shock locations
is not shown there. It should be reminded that a different shock
location $r_{sh}$ is accompanied by a different angular momentum
$\ell$ in this figure. We will discuss this later in \S{3.3}.

\begin{figure}
    \epsscale{0.7}
    \plotone{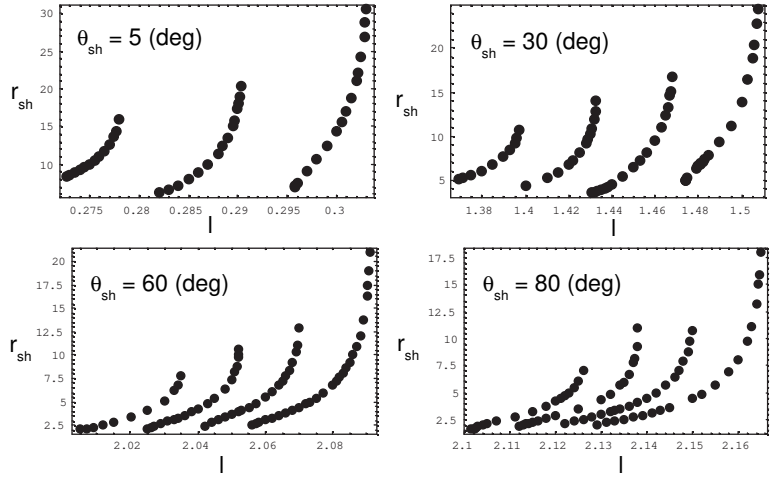}
    \caption{The shock location $r_{sh}$ versus angular momentum $\ell$ for various $E_1$ and $\theta_{sh}$
             in the case of $a=0.99$ (with prograde flows). The dots
             represent the obtained shock locations.
             Each curve corresponds to the respective curves in
             Figure~\ref{fig:shock-location}. From top to bottom, $E_1=1.005,~1.01,~1.015$ and $1.02$ except for
             $\theta_{sh}=5\degr$ where no shock is found for $E_1=1.02$. }
    \label{fig:angular-momentum}
\end{figure}

\placefigure{fig:angular-momentum}

We find that the non-equatorial shock develops further away as
$\ell$ increases for a fixed $E_1$ and $\theta_{sh}$, consistent
with LY98 in the equatorial case. Both the inner critical point
$r^{in}_c$ and the outer critical point $r^{out}_c$ will shift
inward as $\ell$ increases for a fixed $E_1$ and $\theta_{sh}$.
There exist a minimum $\ell$ ($\ell_{min}$) and a maximum $\ell$
($\ell_{max}$) for each set of $(E_1,\theta_{sh})$ for which the
physically valid, stable shock formation is possible. In other
words, the shock is unable to develop because of the violation of
the boundary conditions and/or because of the instability under
ceratin $\ell$ with $(E_1,\theta_{sh})$. This is also commonly
found in the equatorial shocks. The relation between the shock
location $r_{sh}$ and the angular momentum $\ell$ for $a=0.99$ is
seen in Figure~\ref{fig:angular-momentum} where $r_{sh}$ is
plotted against $\ell$ for various angle $\theta_{sh}$.
$E_1=1.005,~1.01,~1.015$ and $1.02$ from top to bottom except for
$\theta_{sh}=5\degr$ where $E_1=1.005,~1.01$ and $1.015$. Notice
that infinitely many shocks do continuously exist between any two
consecutive shocks plotted here although only representative
shocks are shown (in dots) in this figure.

The range of the non-equatorial shock locations $\Delta r_{sh}$ in
general tends to become narrower as $E_1$ increases for a fixed
$\theta_{sh}$, where $\Delta r_{sh} \equiv r^{max}_{sh} -
r^{min}_{sh}$. $r^{min}_{sh}$ and $r^{max}_{sh}$ are the innermost
(smallest) shock location and the outermost (largest) shock
location in the inner branch, respectively. From
Figure~\ref{fig:shock-location}, such a trend is obvious. For a
fixed angle $\theta_{sh}$, the average location of the inner
critical point $\bar{r}^{in}_c \equiv
(r^{in}_{c,min}+r^{in}_{c,max})/2$ seems to be larger while the
average location of the outer point $\bar{r}^{out}_c \equiv
(r^{out}_{c,min}+r^{out}_{c,max})/2$ seems to be smaller as $E_1$
increases, where the additional subscript ``min'' and ``max''
denote the minimum value and the maximum value of the critical
points for a fixed $E_1$ and $\theta_{sh}$, respectively. Another
interesting feature with regard to the angular dependence is that
$\Delta r_{sh}$ roughly becomes larger with decreasing
$\theta_{sh}$ for a fixed $E_1$. That is, both the innermost shock
location $r^{min}_{sh}$ and the outermost shock location
$r^{max}_{sh}$ occur farther away from the hole when the accreting
flow is nearer the polar region although $r^{max}_{sh}$ shifts
outwards more than $r^{min}_{sh}$. This is because the average
$r^{in}_c$ and the average $r^{out}_c$ both tend to become smaller
with increasing $\theta_{sh}$. However, the degree of decrease in
$r^{out}_c$ is much larger than that of $r^{in}_c$, which forces
$\Delta_{sh}$ to be smaller with larger $\theta_{sh}$.
Interestingly, no shocks can form when $E_1=1.02$ with
$\theta_{sh}=5\degr$ as seen in (d). This is due to the presence
of the outer critical points very close to the black hole. With
such a small $r^{out}_c$ (together with a certain value of
$r^{in}_c$), there can be only a narrow (radial) region between
$r^{in}_c$ and $r^{out}_c$. In the case of $5\degr$ in (d), this
region becomes so narrow that no jump conditions can be met within
this range, resulting in no shock formations at all regardless of
the angular momentum $\ell$.

As discussed in the earlier section \S{2.4}, the average shock
location seems to shift outwards with decreasing $\theta_{sh}$,
which is qualitatively predicted as a result of a quick gain of
the radial velocity due to smaller angular momentum $\ell$. In
general, therefore, if the total preshock energy $E_1$ is
relatively large when $\theta_{sh}$ is small, the formation of
shocks may not necessarily be expected. In other words, the
formation of shocks in a polar region may require less energetic
hydrodynamic accretion fluid. The characteristic transition of
these important locations are summarized in Table~\ref{trend}.

\placetable{tab:trend}

\subsection{Properties of Dissipative Shocks }
\begin{figure}
    \epsscale{0.6}
    \plotone{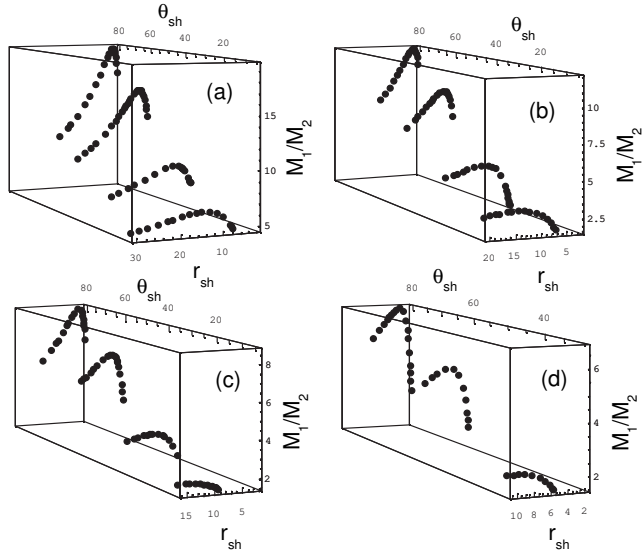}
    \caption{Shock strength (Mach number ratio) $M_1/M_2$ as a function of the shock location $r_{sh}$ and
             the angle $\theta_{sh}$ for the same shocks as in Figure~\ref{fig:shock-location} ($r^{min}_{sh} \le r_{sh}
             \le r^{max}_{sh}$). The dots represent the obtained shocks. $E=1.005,~1.01,~1.015$ and $1.02$
             for (a), (b), (c) and (d), respectively.
             We take $a=0.99$ with prograde flows.}
    \label{fig:mach}
\end{figure}

Figure~\ref{fig:mach} shows a three-dimensional plot where the
shock location $r_{sh}$, the shocked fluid angle $\theta_{sh}$ and
the shock strength $M_1/M_2$ are displayed for various $E_1$. We
take $a=0.99$ (for the prograde flows) with $E_1=1.005, ~1.01,
~1.015$ and $1.02$ for (a), (b), (c) and (d), respectively. Here,
$M_1/M_2$ is the fluid Mach number ratio just before the shock and
just after the shock. For clarity purpose, the range of the plot
varies from (a) to (d). Once again, no shock is found for
$\theta_{sh}=5\degr$ when $E_1=1.02$ (see panel (d)). For a global
geometry of the whole shock location, refer to
Figure~\ref{fig:shock-location}.

\placefigure{fig:mach}

One of the easily noticeable features in this figure is that
$M_1/M_2$ increases with decreasing $r_{sh}$ down to a certain
distance (peak radius $r_p$) and then turns to decrease inside
$r_p$ for a fixed $E_1$. This is again the same pattern found in
LY98 for $\theta$ = $90\degr$ case. Such a trend can be seen more
explicitly as $\theta_{sh}$ increases (compare $5\degr$ and
$80\degr$, for instance). However, as $E_1$ increases, the rate at
which $M_1/M_2$ increases outside $r_p$ seems to become smaller
while the rate at which $M_1/M_2$ decreases inside $r_p$ appears
to be larger for a fixed $\theta_{sh}$. That is, the slope of the
$M_1/M_2-r_{sh}$ curve tends to become smaller with increasing
$E_1$ (compare (a)1.005 and (d)1.02, for instance). Therefore, we
can conclude that the preshock fluid with a small (large) $E_1$
and a large (small) $\theta_{sh}$ is more likely to produce a
strong (weak) shock for a fixed shock location $r_{sh}$.

\begin{figure}
    \epsscale{0.6}
    \plotone{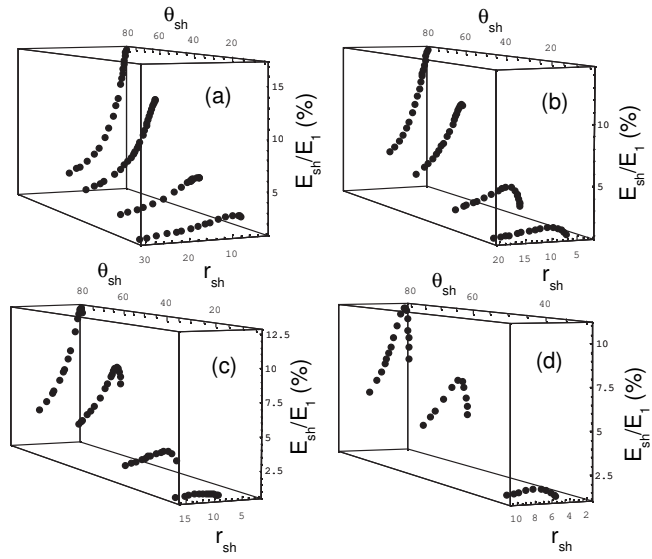}
    \caption{The ratio of the energy dissipation to the preshock energy $E_{sh}/E_1$ (in \%) for the same
             shocks as in Figure~\ref{fig:shock-location}. }
    \label{fig:energy}
\end{figure}

\placefigure{fig:energy}

The energy dissipation $E_{sh}/E_1$ (or the ratio of the energy
released to the preshock energy) from the preshock fluid is shown
(in percent) as a function of $r_{sh}$ and $\theta_{sh}$ in
Figures~\ref{fig:energy}. The panels (a)-(d) correspond to the
same energy $E_1$ as in Figure~\ref{fig:mach}. This ratio
basically follows the same pattern as the $M_1/M_2$ except for a
specific detail. Since the $E_{sh}/E_1-r_{sh}$ curve qualitatively
resembles the $M_1/M_2-r_{sh}$ curve, the energy dissipation
$E_{sh}$ becomes larger as the shock occurs closer to the hole
until the peak radius $r_p$ if it exists at all. The release of
the energy from the hot flow tends to be more efficient when
$\theta_{sh}$ is large for a fixed energy $E_1$, and the variation
of $E_{sh}/E_1$ can be nearly an order of magnitude if the shock
is developed near the disk plane (i.e., $\theta_{sh} \sim
90\degr$) which can be seen in (a) and (b) in
Figure~\ref{fig:energy}. This can be qualitatively interpreted in
terms of the energy conversion of the fluid, as we predicted in
the section \S{2.4}. The Shock developed further away can
dissipate only a small gravitational potential energy whereas the
shock close to the hole can convert more potential energy for
dissipation. The maximum energy release can be as high as $\sim
18\%$ whereas the minimum energy release seems less than $1\%$ in
the case of $a=0.99$ (with prograde flows). As another trend, more
energy $E_{sh}$ tends to be released from a hot flow with a
smaller $E_1$. Overall, it can be concluded that more energy
release is expected from a low-latitude (i.e., large
$\theta_{sh}$) hot flow with a small energy $E_1$.

\placetable{tab:para1}
\placetable{tab:flow}

\subsection{Angular Dependence}
\begin{figure}
    \epsscale{0.5}
    \plotone{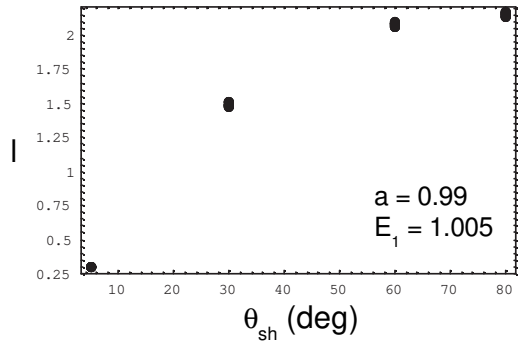}
    \caption{The specific angular momentum of the flow $\ell$ versus the angle $\theta_{sh}$. We take
             $\theta_{sh}=5\degr,~30\degr,~60\degr$ and
             $80\degr$. Each dot is a collection of many smaller
             dots that correspond to particular angular momentum
             at different radial shock locations $r_{sh}$. The other parameters
             are $a=0.99$ and $E_1=1.005$.}
    \label{fig:angle-ell}
\end{figure}

In order to see the angular dependence of the specific angular
momentum of the shocked fluid, $\ell$, Figure~\ref{fig:angle-ell}
shows the $\ell$ versus $\theta_{sh}$ relation. We choose preshock
fluid energy and black hole spin at fixed values, $E_1=1.005$ and
$a=0.99$, respectively, in order to better understand the effects
due to $\theta_{sh}$ alone.  We find that at each $\theta_{sh}$,
$\ell$ ranges from $\ell_{min}$ to $\ell_{max}$ for different
radial shock location $r_{sh}$, (although it may be difficult to
see this in the figure because $\ell_{mx}-\ell_{min}$ is much
smaller than the variation of $\ell$ over $\theta_{sh}$). One can
express the angular momentum as a function of $r-{sh}$, as $\ell
=\ell(r_{sh})$. Note that LY98 also finds similar behavior (i.e.,
the $r_{sh}$ dependence of $\ell$). In fact, this is a general
behavior found for any other shock-included global solutions.  On
the other hand, we note that for shock-free global accreting flows
the $r$ dependence disappears, and hence a single value of $\ell$
should be determined by specifying $\theta$ alone.

We clearly see (in Figure~\ref{fig:angle-ell}) the general trend
that the fluid angular momentum decreases with decreasing
$\theta_{sh}$. That is physically consistent with the fact that
the centrifugal force (due to the gravitational potential well)
decreases towards the polar axis. Thus, the fluid must possess
smaller angular momentum in the polar region (i.e., for small
$\theta_{sh}$) for accretion to be realized. A similar trend has
been found also in the work of the MHD shocks by TRFT02. It may be
possible, in principle, to analytically show the dependence of the
physically allowed angular momentum $\ell$ as a function of
$\theta_{sh}$, for a given value of $r_{sh}$ (i.e.,
$\ell(r_{sh},\theta_{sh})$). However, that is beyond the scope of
our present work.  For our computation purposes these two
variables ($\ell(r_{sh})$ and $\theta_{sh}$) are, to start with,
treated as independent of each other. However, the valid solutions
numerically found do indicate that $\ell$ is uniquely determined
for a given set of the shock location ($r_{sh}, \theta_{sh}$).

\placefigure{fig:angle-ell}

\subsection{Black Hole Spin Dependance}
\begin{figure}
    \epsscale{0.5}
    \plotone{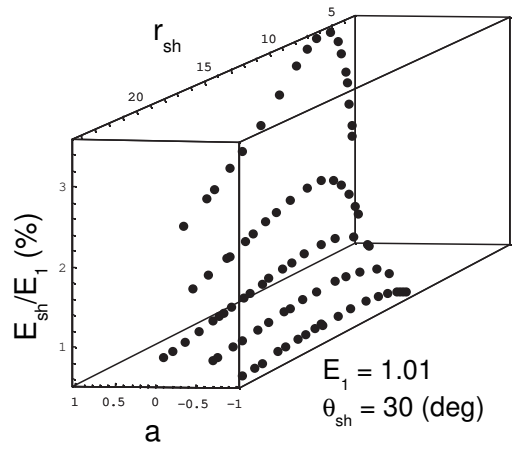}
    \caption{The energy dissipation ratio $E_{sh}/E_1$ (in \%) as a function of the shock location $r_{sh}$
             and the black hole spin $a$ for $E_1=1.01$ and
             $\theta_{sh}=30\degr$. The dots represent the
             obtained shocks.
             We choose $a=-0.99,~-0.5,~0,~0.5$ and $0.99$.
             Negative black hole spin cases are considered to have retrograde flows (i.e., $a \ell <0$).  }
    \label{fig:spin-dependence}
\end{figure}

Here we explore the effect of black hole spin on shock properties.
To decouple the effect of the hole spin from the other effects, we
hold the fluid energy and the angle at fixed values of $E_1=1.01$
and $\theta_{sh}=30\degr$, respectively.
Figure~\ref{fig:spin-dependence} demonstrates the energy
dissipation $E_{sh}/E_1$ (in \%) as a function of the spin
parameter $a$ and shock location $r_{sh}$. We take
$a=-0.99,~-0.5,~0,~0.5$ and $0.99$. Since the fluid angular
momentum $\ell$ is always taken to be positive here, the cases
$a=-0.99$ and $-0.5$ refer to retrograde flows.

First we note that energy is most efficiently dissipated around a
rapidly-rotating black hole in the presence of prograde flows. The
retrograde flows with $a=-0.99$, on the other hand, appear to
produce the least energy release, which corresponds to small Mach
number ratio $M_1/M_2$. Due to the frame-dragging by the hole, the
prograde flows can quickly acquire more velocity (especially the
rotational component) in the course of its accretion, as opposed
to the retrograde case where the fluid rotation appears even
smaller (in the local inertial frame). Therefore, the prograde
flows can afford more change in the kinetic motion at a shock
location, and thus it may be ``easier'' for the prograde flows to
dissipate more energy using the obtained kinetic motion. This
viewpoint can be directly translated into the large Mach number
ratio $M_1/M_2$ that we generally observe in the case of the
prograde accretion. This may naturally explain large energy
release $E_{sh}/E_1$ at shocks that we normally find. On the
contrary, the retrograde flows generally cannot afford as much
change as the prograde can in terms of the kinetic motion, and
therefore only weak shocks can develop with small Mach number
ratio (or small energy dissipation).

Keeping in mind that we are only interested in the inner range of
the shocks here, the minimum shock location $r^{min}_{sh}$ tends
to shift radially outwards with decreasing spin $a$. The maximum
shock location $r^{max}_{sh}$ is the greatest when $a=0$. This
suggests that in the retrograde flows no shock formations should
be expected in a region very close to the hole, as opposed to the
prograde case.

We can conclude from Figure~\ref{fig:spin-dependence} that very
strong shocks (with more energy release) are expected in the inner
region for the prograde flows.

\placefigure{fig:spin-dependence}

\subsection{Global Shock-Included Accreting Flow Solutions }
\begin{figure}
    \epsscale{0.5}
    \plotone{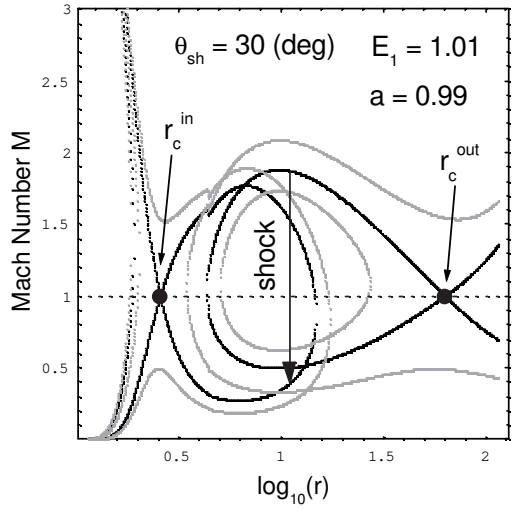}
    \caption{A flow topology of a global shock-included accreting flow with $E_1=1.01$ and $\theta_{sh}=30\degr$ for $a=0.99$.
             A preshock transonic flow
             with $(E_1,~\dot{\cal{M}}_1)$ passing through an outer critical point $r^{out}_c$ develops an isothermal shock at
             $r_{sh}$ and becomes a subsonic postshock flow
             with $(E_2,~\dot{\cal{M}}_2)$. The flow then passes through an inner critical point
             $r^{in}_c$ becoming supersonic again
             before reaching the horizon. The physically valid preshock/postshock flows
             through the critical points are denoted by {\it dark curves},
             while {\it light grey curves} refer to unphysical flows.}
    \label{fig:topology}
\end{figure}

As explained in Section \S{2.3}, in the case of isothermal shocks,
the global shock-included transonic solution must pass through the
outer critical (sonic) point determined by the preshock parameters
and simultaneously goes through the inner critical (sonic) point
determined by the postshock parameters. Before looking into the
detailed of particular flow dynamics, let us illustrate such a
flow topology for a specific example. Figure~\ref{fig:topology}
shows the Mach number $M$ of the transonic flow as a function of
the radial distance $r$ (in logarithmic scale) for a
shock-included transonic solution with $a=0.99,~\ell=1.463$ and
$\theta_{sh}=30\degr$. The preshock flow with
$(E_1,\dot{\cal{M}}_1)=(1.01,5.27 \times 10^{-5})$ goes through
the outer critical point at $r^{out}_c = 63.2 r_g$, then makes the
transition via the shock at $r_{sh}=11 r_g$ into the postshock
flow with $(E_2,\dot{\cal{M}}_2)=(0.98258,1.06 \times 10^{-5})$
passing through the inner critical point at $r^{in}_c = 2.58 r_g$
before reaching the horizon. Notice that the preshock flow
topology is completely independent of the postshock flow topology
(both drawn in dark curves) in that they individually possess
their own critical points (i.e., non-mutual points), and the flow
transition from the former to the latter is allowed only through
the shock (denoted in arrow). The other flows (shown in light gray
curves) with different values of $\dot{\cal{M}}$ are unphysical
solutions in terms of the fact that the jump conditions are not
satisfied and that no critical points exist.

\placefigure{fig:topology}

\begin{figure}
    \epsscale{0.85}
    \plotone{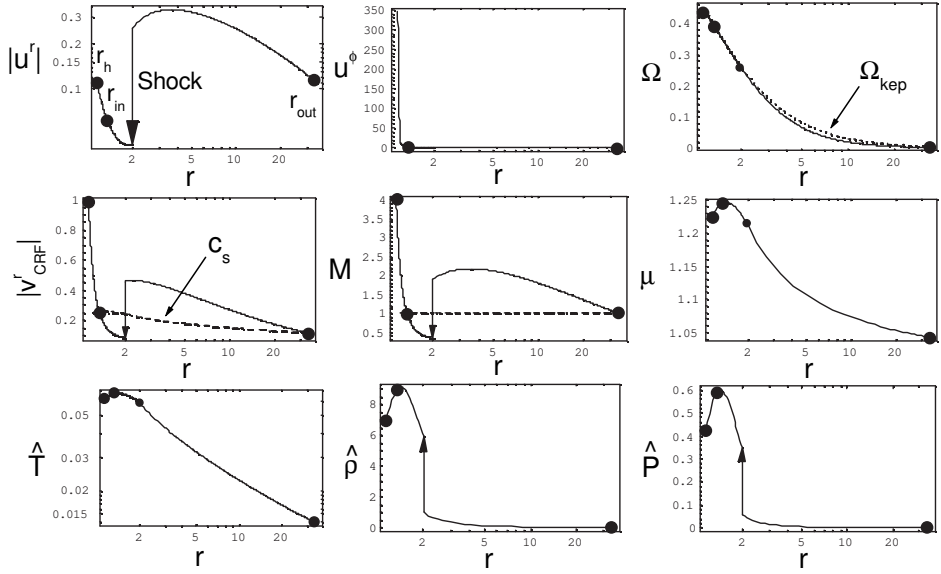}
    \caption{Global shock solutions for an adiabatic fluid (flow 1) with $E_1=1.02,\ell=2.104,r_{sh}=2.005r_g$
             for $\theta_{sh}=80\degr$.
             The horizon, inner and outer critical (sonic) points are respectively denoted by
             {\it large
             dots} from left to right in each figure.
             The shock is represented by either a {\it small dot} or a {\it vertical arrow}. From the upper-left
             panel to the upper-right panel,
             the radial component $|u^r|$ of the flow, the azimuthal component $u^{\phi}$ of the flow and the angular velocity
             $\Omega$ ({\it solid curve}) with the Keplerian value $\Omega_{kep}$ ({\it dotted curve}) are shown.
             From the middle-left panel to the middle-right panel, the radial three-velocity $|v^r|$ of the flow in the
             corotating reference frame (CRF) ({\it solid curve}) with the local sound velocity $c_s$ ({\it dotted curve}),
             the Mach number $M$ and the relativistic enthalpy
             $\mu$ are displayed. From the lower-left panel to the lower-right panel, the fluid local temperature $\hat{T}$,
             the density $\hat{\rho}$, and the thermal pressure of the flow $\hat{P}$ are plotted.
             For details, see Tables~\ref{para1}
             and \ref{flow}.}
    \label{fig:global-soln1}
\end{figure}

\begin{figure}
    \epsscale{0.85}
    \plotone{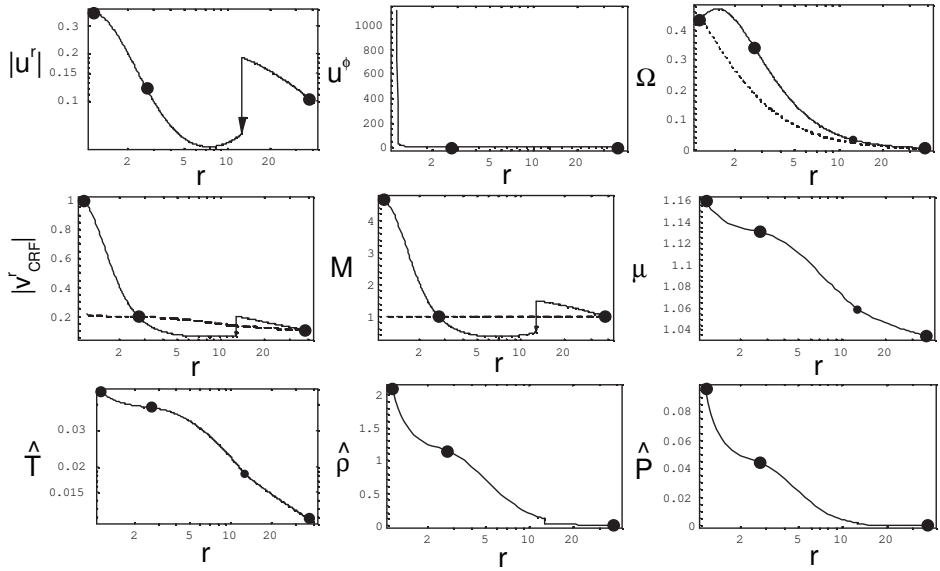}
    \caption{Global shock solutions for an adiabatic fluid (flow 2) with $E_1=1.015,\ell=1.4325,r_{sh}=12.75r_g$
             for $\theta_{sh}=30\degr$. The notations are the same as in Figure~\ref{fig:global-soln1}.}
    \label{fig:global-soln2}
\end{figure}

\begin{figure}
    \epsscale{0.85}
    \plotone{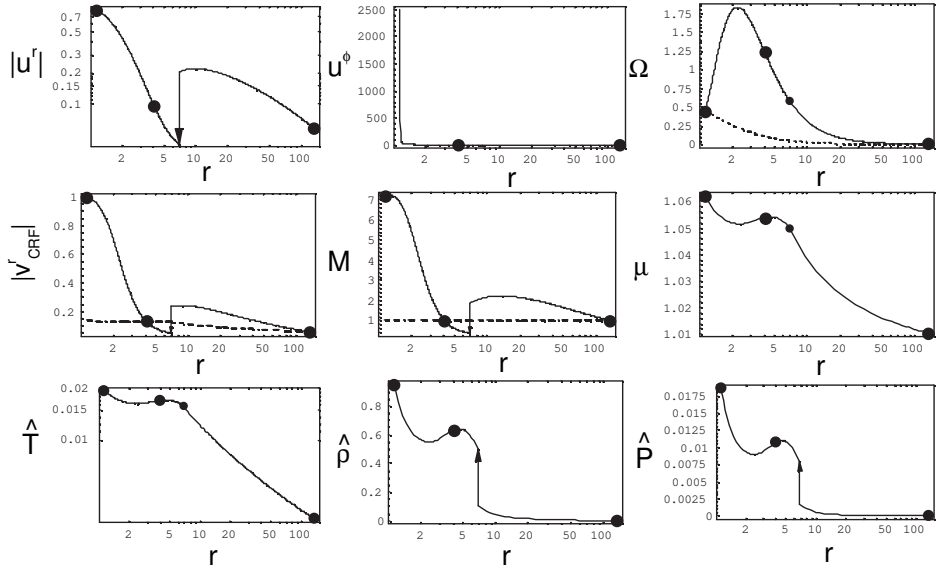}
    \caption{Global shock solutions for an adiabatic fluid (flow 3) with $E_1=1.005,\ell=0.2958,r_{sh}=6.95r_g$
             for $\theta_{sh}=5\degr$. The notations are the same as in Figure~\ref{fig:global-soln1}.}
    \label{fig:global-soln3}
\end{figure}

Here, we present three types of global, shock-included, transonic,
accreting flow solutions in the case of $a=0.99$ (with prograde
flows): Figure~\ref{fig:global-soln1} for flow 1,
Figure~\ref{fig:global-soln2} for flow 2 and
Figure~\ref{fig:global-soln3} for flow 3. The flow parameters are
tabulated in Table~\ref{para1} for each flow. Each figure displays
the radial component $|u^r|$ and the azimuthal component
$u^{\phi}$ of the four velocity, the angular velocity $\Omega$
(solid curve) with the equatorial Keplerian velocity
$\Omega_{kep}$ (dotted curve), the radial three-velocity in CRF
$|v^r|$ (solid curve) with the local sound velocity $c_s$ (dotted
curve), Mach number $M$, specific enthalpy $\mu$, flow temperature
$\hat{T}$, local density $\hat{\rho}$ and thermal pressure
$\hat{P}$ respectively. The enthalpy here is also normalized by
the fluid rest-mass energy $m_p c^2$. The outer/inner critical
(sonic) point is denoted by a large dot. The value at the event
horizon is shown by another large dot. Either a vertical arrow or
a small dot shows the transition from a preshock flow to a
postshock flow through a stable isothermal shock. Note that
$r_{hor} < r^{in}_c < r_{sh} < r^{out}_c$ for physically
acceptable shocks. In Table~\ref{flow}, some representative fluid
quantities for each flow is displayed in physical units. To make
concrete calculations, we adopt the following relation between the
(dimensionless) mass accretion rate for the non-equatorial hot gas
($\dot{m}_{hot}$) and the observed net accretion rate
($\dot{m}_{net}$) in the units of the Eddington mass accretion
rate ($\sim 1.4 \times 10^{25}$ g/s) for a $10^7 M_{\sun}$ black
hole mass. Here, we take $\dot{m}_{hot} \sim 0.01 \dot{m}_{net}$
and $\dot{m}_{net}=0.1$ (relevant for radio-quiet AGNs) based on
the assumption that the most of the accretion rate (99\%) is
contributed from the cool accreting gas in the accretion disk,
whereas only a fraction of the net accretion rate ($\sim$ 1\%) is
due to the hot non-equatorial accreting fluid.

\placefigure{fig:global-soln1}

As seen in the earlier sections, flow 1 in
Figure~\ref{fig:global-soln1} goes through a relatively strong
shock ($M_1/M_2 \sim 5.4$) as a consequence of the dramatic
increase in the density $\hat{\rho}$ and the pressure $\hat{P}$.
The flow temperature $\hat{T}$ seems to be monotonically rising up
to a peak point $r_p$ where the local sound speed is the maximum.
In fact, all the thermodynamic quantities, such as $\mu$,
$\hat{T}$ and $\hat{\rho}$, are all correlated to the sound speed.
Notice that the local sound speed $c_s$ is continuous across the
isothermal shocks because the flow temperature $\hat{T}$ is by
definition continuous. Passing through $r_p$, the sound velocity
starts decreasing which triggers a relatively large decrease in
the thermodynamic properties such as $\mu$, $\hat{\rho}$,
$\hat{P}$ and $\hat{T}$. The angular velocity $\Omega \equiv
u^{\phi}/u^t$ roughly follows the Keplerian value (in the equator)
almost all the way down to the horizon. The radial three-velocity
$|v^r|$ appears to become maximum just before the shock and then
transits through a strong shock, rising up again rapidly to become
supersonic before entering the horizon. It is seen that, as
expected, $|v^r|=1$ at the horizon.

\placefigure{fig:global-soln2}

Flow 2 in Figure~\ref{fig:global-soln2} shows somewhat different
behaviors compared to flow 1. First, the shock strength is weaker
($M_1/M_2 \sim 3.0$). In terms of the kinematics, the angular
velocity reaches a maximum value at some point and then slows down
in the azimuthal direction from this point on. Notice that
$\Omega(r)>\Omega_k$ for its entire trajectory. An interesting
difference here is that the radial velocity is still decreasing
after the shock for about $\sim 5r_g$ and then turns to pick up
the (radial) speed. This is probably due to a relatively large
shock location ($r_{sh} \sim 13r_g$). Thermodynamic quantities
also behave in different ways here. After the shock, the flow
temperature does not drop in this case. Instead, it continues to
rise gradually just after the shock and the rate of the change
becomes larger towards the horizon. This behavior is originating
from the sound speed $c_s(r)$. The rest of the thermodynamic
quantities ($\hat{\rho},~\mu$ and $\hat{P}$) also follow the same
pattern.

\placefigure{fig:global-soln3}

The shock strength for flow 3 in Figure~\ref{fig:global-soln3} is
somewhere between flow 1 and flow 2 ($M_1/M_2 \sim 4.8$). An
outstanding feature in this case is a large deviation of
$\Omega(r)$ from the Keplerian angular velocity in the equator.
Since the shock location is relatively close to the hole, the
postshock (radial) speed increases immediately after the shock. In
both temperature and density profiles, a strange behavior can be
seen after the shock. In temperature profile, the postshock fluid
temperature appears almost constant producing a plateau-like
pattern, whereas density profile $\hat{\rho}$ shows a step-like
change in $\hat{\rho}$. Again, such a feature of these
thermodynamic quantities can be explained by the sound speed
$c_s(r)$.

It appears in general from the above three cases that the
postshock fluid does not become supersonic right away if the shock
is formed at a relatively distant location (say, $r_{sh} \gtrsim
10r_g$), in which case the radial velocity can be still slowing
down. It turns out, however, that it is not easy to classify all
the possible shock-included flows according to the angle
$\theta_{sh}$ because the flow energy $E_1$ as well as the angular
momentum $\ell$ are all certainly related to the flow dynamics,
which allows a varieties of flow dynamics even for a fixed
$\theta_{sh}$. For this reason, we will not try to simply
attribute the above hydrodynamic/thermodynamic features in each
flow to just one parameter out of ($E_1,\ell,\theta_{sh}$).

\section{Discussion \& Concluding Remarks}

Although our current work is partially motivated by the work of
LY98, it also originates, in important ways, from our very recent
work on black hole magnetospheres in accretion-powered AGNs -
TRFT02 and Rilett et al. (2004, in preparation), where we explored
adiabatic, relativistic MHD shocks produced in the black hole
magnetosphere. We showed that strong shocks can indeed be formed
in such flows for various relevant choices of flow parameters. In
the current paper we extended these previous studies to explore
the relativistic hydrodynamic flows which should apply to the case
of weak magnetization. The reason is that the physics involved in
the exact MHD case would be far more complicated, as noted already
by TRFT02. Owing to these complexities, it was not straightforward
for us to study the exact global shock-included MHD accretion flow
solutions further in detail in a wider parameter/solution space.

As our next step, in the current paper we assumed a simple model
of conical accretion flows, due to the fact that modelling more
realistic flows (such as the ones in hydrostatic equilibrium)
would be very complicated, especially in the framework of the
relativistic non-equatorial accretions considered here. An
appropriate force balance under the general relativistic geometry
should be taken into account in more sophisticated models, but
that is beyond the scope of our present work. We, however, would
like to stress that our results still represents (at least
qualitatively) important physical characteristics of shock
formation in non-equatorial accretion flows.  Also, in the
presence of the magnetosphere, the fluid particles would be
frozen-in to the field and hence flow along the field lines. For
such a situation, application of conventional thick accretion disk
(or torus) models is not appropriate.

We considered only the inner range of the shock formations in
regions relatively close to the central engine, because of our
interests in some X-ray observations of the reprocessed emission,
such as the iron fluorescence lines, from some AGNs. Our results
may offer the possibility of a high energy source in various broad
regions ($0\degr < \theta_{sh} < 90\degr$) above the disk plane.
However, we find that the strongest shocks (high $M_1/M_2$ with
large $E_{sh}$) should develop near the equator (large
$\theta_{sh}$), although the quasi-polar shock (small
$\theta_{sh}$) is also possible. We find no shocks in the polar
region ($\theta_{sh}=5\degr$) when the preshock fluid energy is
relatively large although the shock-free accretion is physically
allowed. The magnitude of the energy release from the shock
roughly increases as the shock location gets closer towards the
black hole, as already found by LY98 by their 1D studies. In
addition, however, we further find, from our 2D calculations, that
the shock-induced energy release $E_{sh}$ greatly depends, not
only on the fluid energy $E_1$ and angular momentum $\ell$, but
also on {\it the angle of the shock location $\theta_{sh}$}.  The
average shock strength (thus energy release) tends to be weaker
towards the polar region.

Although TRFT02 considered adiabatic shocks and hence no energy
dissipation, in the current paper we adopted the isothermal shocks
because that ensures a substantial amount of energy release at the
shock locations. Also, through the stability analysis our energy
source (i.e., the shock) is found to be stable. Although we
adopted adiabatic flows in our current work, it may be noted that
\cite{DPM03} recently explored the isothermal shock formations in
the isothermal flows, by adopting pseudo-Schwarzschild
gravitational potentials and 1D equatorial flows. These authors
concluded that their shocks also will release substantial amount
of energy, which could be physically sufficient to become a
radiation source for a strong X-ray flare.

In the present work, it is argued that a single isothermal shock
formation between the two sets of accreting flows
(preshock/postshock flows) is very likely with a substantial
amount of energy dissipation. One may also ask whether it is
possible to have a sequence of shock formations one after another
in the course of the accretion. For example, a preshock flow with
energy $E_1$ gets shocked at a shock location $r_{sh,1}$ becoming
a subsonic postshock flow with energy $E_2$. The same flow with
energy $E_2$ then becomes supersonic and develops another shock at
$r_{sh,2}$ where $r_{sh,2}<r_{sh,1}$, making the transition to
another subsonic postshock flow with energy $E_3$ where
$E_3<E_2<E_1$. Such a sequential shock formation (namely ``shock
cascade'' or ``shock avalanche'') could be a very interesting
phenomenon. Although discussion of such a possibility is beyond
the scope of the present work, it is interesting, in a future
work, to investigate such a possibility in relation to some
observational applicability.

Before closing, we emphasize that a major justification for the
choice of our version of a conical flow geometry follows from our
recent work of TRFT02 on magnetospheric MHD accretion flows.
However, here we adopted the hydrodynamic flows, as a limiting
case of weak magnetization, in order to take advantage of the
current fortunate situation that the relativistic hydrodynamic
(equatorial) shocks in 1D accreting flows have already been
extensively studied by many authors \cite[e.g.,][and
LY98]{Cha89,Abramowicz90,Lu97a,Lu97b}. Our current investigation,
however, is new and valuable in the sense that {\it we have
explored 2D non-equatorial shocks in a fully relativistic manner,
with a possible application, for instance, to offer an attractive
definite source for the energy dissipation in regions very close
to the black hole.}

\acknowledgments

We are grateful to Darrell Rilett and Masaaki Takahashi for
enlightening suggestions on our model. KF also thanks Maki
Fukumura for computational assistance. ST thanks colleagues in
Institute of Astronomy, Cambridge, especially Drs. A. Fabian, G.
Miniutti, M.J. Rees, and K. Iwasawa, for valuable discussions and
comments. We are especially indebted to the anonymous referee for
providing a number of constructive suggestions to improve the
manuscript.

\clearpage

\clearpage

\clearpage

\clearpage

\clearpage

\clearpage

\clearpage

\clearpage

\clearpage

\clearpage

\clearpage

\clearpage

\begin{deluxetable}{ccrrrrr}
\tabletypesize{\scriptsize} \tablecaption{Some Important Locations
as a Function of $E_1$ and $\theta_{sh}$. \label{trend}}
\tablewidth{0pt} \tablehead{ \colhead{$E_1$}&
\colhead{$\theta_{sh}$}  & \colhead{$r^{min}_{sh}$} &
\colhead{$r^{max}_{sh}$} & \colhead{$\Delta r_{sh}$} &
\colhead{$\bar{r}^{in}_c$} & \colhead{$\bar{r}^{out}_c$} }
\startdata
1.005 & $5\degr$ & 6.950 & 30.50 & 23.55 & 3.868 & 132.7  \\
1.005 & $60\degr$ & 4.970 & 21.00 & 16.03 & 1.377 & 144.3  \\
1.005 & $80\degr$ & 3.625 & 18.00 & 14.38 & 1.262 & 146.0  \\
1.01 & $5\degr$ & 6.188 & 20.32 & 14.13 & 4.238 & 57.71   \\
1.01 & $60\degr$ & 4.050 & 12.85 & 8.810 & 1.420 & 69.28  \\
1.01 & $80\degr$ & 3.330 & 10.80 & 7.471 & 1.278 & 71.05  \\
1.02 & $30\degr$ & 5.062 & 10.62 & 5.559 & 3.048 & 25.62  \\
1.02 & $60\degr$ & 2.060 & 7.800 & 5.741 & 1.515 & 31.77  \\
1.02 & $80\degr$ & 1.650 & 7.000 & 5.352 & 1.316 & 33.56  \\
\enddata
\tablecomments{The length (or distance) is in the unit of $r_g$.
See text for notations. $a=0.99$ for all cases.}
\end{deluxetable}

\clearpage

\begin{deluxetable}{crrrrrrrrrr}
\tabletypesize{\scriptsize} \tablecaption{Global Shock-Included,
Transonic, Non-Equatorial Accretions for Various Flow Parameters.
\label{para1}} \tablewidth{0pt} \tablehead{ \colhead{Flow}&
\colhead{$E_1$}  & \colhead{$\ell$} & \colhead{$\theta_{sh}$} &
\colhead{$r_{sh}/r_g$} & \colhead{$r^{in}_c/r_g$} &
\colhead{$r^{out}_c/r_g$} & \colhead{$M_1/M_2$} &
\colhead{$\dot{\cal{M}}_1(10^{-5})$} &
\colhead{$\dot{\cal{M}}_2(10^{-5})$} & \colhead{Figure} }
\startdata
1 & 1.02 & 2.104 & $80\degr$ & 2.005 & 1.347 & 33.65 & 5.398 & 15.43 & 2.551  & \ref{fig:global-soln1} \\
2 & 1.015 & 1.4325 & $30\degr$ & 12.75 & 2.697 & 37.85 & 2.987 & 9.265 & 3.046  & \ref{fig:global-soln2} \\
3 & 1.005 & 0.2958 & $5\degr$ & 6.95 & 4.059 & 133.3 & 4.761 & 1.906 & 0.3896 & \ref{fig:global-soln3} \\
\enddata
\tablecomments{$a=0.99$ for all cases. The black hole mass $M$ is
$10^7 M_{\sun}$.}
\end{deluxetable}

\clearpage

\begin{deluxetable}{crrrrrrrrr}
\tabletypesize{\scriptsize} \tablecaption{Thermodynamic Properties
of Accreting Hot Flows at the Shock Location. \label{flow}}
\tablewidth{0pt} \tablehead{ \colhead{Flow} & \colhead{$E_1$}  &
\colhead{$\theta_{sh}$} & \colhead{$r_{sh}/r_g$} &
\colhead{$c_s/c$} & \colhead{$n_1(10^{10}$)\tablenotemark{a}} &
\colhead{$n_2(10^{10})$\tablenotemark{a}} &
\colhead{$P_1(10^5)$\tablenotemark{b}} &
\colhead{$P_2(10^5)$\tablenotemark{b}} & \colhead{Figure}}
\startdata
1 & 1.02 & $80\degr$ & 2.005 & 0.2420 & 12.06 & 73.00 & 79.67 & 482.0 & \ref{fig:global-soln1} \\
2 & 1.015 & $30\degr$ & 12.75 & 0.1351 & 0.4051 & 1.232 & 0.8335 & 2.535 & \ref{fig:global-soln2}  \\
3 & 1.005  & $5\degr$ & 6.95 & 0.1258 & 1.253 & 6.129 & 2.235 & 10.93 & \ref{fig:global-soln3} \\
\enddata
\tablenotetext{a}{in 1/cm$^3$} \tablenotetext{b}{in dyne/cm$^2$}
\tablecomments{The subscripts ``1'' and ``2'' denote the preshock
and postshock quantities, respectively. We adopt black hole mass
$M=10^7 M_{\sun}$, and $\dot{m}_{hot}=0.01 \dot{m}_{net}=10^{-3}$.
See the text for notations.}
\end{deluxetable}

\end{document}